\def\Ar{\rightarrow}
\def\bar{\overline}

\def\n{\nu}

\def\k{\kappa}
\def\e{\epsilon}

\def\bar{\overline}
\def\l{\lambda}
\def\G{{\rm GeV}}

\documentstyle [12pt,epsf]{article}
\begin{document}
\baselineskip=22 pt
\setcounter{page}{1}
\thispagestyle{empty}
\topskip 2.5  cm
%%%%%%%%%%%%%%%%%%%%%%%%%%%%%%%%%%%%%%%%%%%%%%%%%%%%%%%%%%%%%%%%%%%%%%
\setcounter{page}{1}
\thispagestyle{empty}
\topskip 0.5  cm
%\begin{flushright}
%\begin{tabular}{c c}
%& {\normalsize hep-ph/012}\\
%& , March 2002
%\end{tabular}
%\end{flushright}
\vspace{1 cm}
\centerline{\Large\bf Neutrino Mass Matrix 
   with Two Zeros}
\vskip 0.5 cm
\centerline{\Large\bf and  Leptogenesis}
\vskip 1.5 cm
\centerline{{\large \bf Satoru Kaneko}
\renewcommand{\thefootnote}{\fnsymbol{footnote}}
\footnote[1]{E-mail address: kaneko@muse.sc.niigata-u.ac.jp}  
\quad {\large\bf and}
\quad {\large \bf Morimitsu Tanimoto}
\renewcommand{\thefootnote}{\fnsymbol{footnote}}
\footnote[2]{E-mail address: tanimoto@muse.sc.niigata-u.ac.jp}
 }
\vskip 0.8 cm
 \centerline{ \it{Department of Physics, Niigata University, 
 Ikarashi 2-8050, 950-2181 Niigata, JAPAN}}
\vskip 4 cm
\centerline{\bf ABSTRACT}\par
\vskip 1 cm
The leptogenesis is studied in the neutrino mass matrix with two zeros, which 
reduces the number of independent phases of the  CP violation. 
These texture zeros are decomposed into the Dirac neutrino mass matrix and
the right-handed Majorana one in the see-saw mechanism.
It is found that there are six textures for the Dirac neutrino mass matrix, 
among  which one texture with three zeros has three phases, three textures
  with four zeros have two phases, and two textures  with five zeros have
only one phase  in the basis with the real right-handed Majorana mass matrix. 
The relation between the leptogenesis and the low energy CP violation 
  is discussed in typical textures. 
It is remarked that the leptogenesis is linked strongly with the CP violation
 in the neutrino oscillations.
The predicted baryon asymmetry of the universe is consistent 
with the observed value.

\newpage
%%%%%%%%%%%%%%%%%%%%%%%%%%%%%%%%%%%%%%%%%%%%%%%%%%%%%%%%%%%%%%%%%%%%%%%%%%%%%%%
%%%%%%%%%%%%%%%%%%%%%%%%%%%%%%%%%%%%%%%%%%%%%%%%%%%%%%%%%%%%%%%%%%%%%%%%%%%%%%%
\topskip 0. cm

 The Super-Kamiokande has almost confirmed the neutrino oscillation
 in the atmospheric neutrinos, which favors the $\n_\mu\Ar \nu_\tau$
process  \cite{SKam}.
For the solar neutrinos \cite{SKamsolar,SNO}, the recent data of 
the Super-Kamiokande and the Sudbury Neutrino Observatory 
 also favor strongly the neutrino oscillation
$\n_e\Ar \nu_{\mu,\tau}$ with the large mixing angle (LMA) MSW solution 
\cite{MSW,Lisi}. 
These results indicate the neutrino masses and mixings \cite{MNS},    
especially, the bi-large flavor mixing. 
It is therefore possible to investigate the textures of the neutrino  mass 
matrix.

 Recently, Frampton, Glashow and Marfatia \cite{Fram} found seven
acceptable textures of the neutrino mass matrix with two independent
vanishing entries in the basis with the diagonal charged lepton mass matrix. 
 The further studies of these textures  
were presented by some authors \cite{Xing,Desai,Yoshikawa,KKST}.  
Another insight has been also given focusing on texture zeros \cite{Kang}.
Since these textures are given for the light left-handed neutrinos,
 one needs to find the see-saw realization \cite{Seesaw}
of these textures from the standpoint of the model building.
We have found  the Dirac neutrino
mass matrices  and the right-handed Majorana neutrino ones, which give
 those seven textures \cite{KKST}.  

These textures should be tested in the future experiments.
We have already shown that they can be tested
in the lepton flavor violating processes, especially, 
$\mu \rightarrow e \gamma$  \cite{KKST}.   
In order to search another test of them,
  we discuss the leptogenesis in this paper.
We study the relation between the leptogenesis and the low energy CP violation
  in these textures.
We also predict the magnitude of the baryon asymmetry of the universe.

The leptogenesis \cite{Fu} is an interesting mechanism to explain
 the observed baryon asymmetry of the universe.
Since the leptogenesis depends strongly on the texture and
 the CP violating phases of the neutrino mass matrix,  it is
expected that the leptogenesis links with the low energy neutrino phenomena,
especially the low energy CP violation  in the neutrino oscillations
 \cite{lepto,leptcp,lepttexture}.
The CP violation required for the leptogenesis stems from the CP violating
phases in the right-handed  sector, whereas
the CP violation in  the neutrino oscillations \cite{CP}
can be described by the left-handed neutrino mixing matrix.
In general, the independent CP violating phases are six
for  three generations and without a left-handed Majorana mass term, 
and therefore
it is very difficult to find links between the leptogenesis and 
the low energy CP violation.  Reduction of the number of phases
has been done by considering the two right-handed neutrinos \cite{FGY,moro}.
The interesting  connection  between the leptogenesis and 
the low energy CP violation has been presented in those works.
 It is remarked that there is one massless neutrino in this scheme.

In order to reduce the number of phases, we propose another approach,  
which is the texture zeros with the three right-handed neutrinos,
being  originated from 
the work by Frampton, Glashow and Marfatia \cite{Fram}. We show that 
the low energy leptonic CP violation can be affected by the CP violating 
phases responsible for the leptogenesis in these texture zeros.

%%%%%%%%%%%%%%%%%%%%%%%%%%%%%%%%%%%%%%%%%%%%%%
%%%  Texture Zeros and Seesaw Mechanism  %%%%%
%%%%%%%%%%%%%%%%%%%%%%%%%%%%%%%%%%%%%%%%%%%%%%%%%%%%%%%%%%%%%%%%%%%%%%%%%%
Let us discuss the structure of the CP violating phases in the neutrino mass
matrix.
 There are seven acceptable textures with two independent zeros
for the effective neutrino mass matrix   $M_{\nu}$ \cite{Fram}.
The two textures ($\rm A_1$ and $\rm A_2$) correspond to the hierarchical
neutrino mass spectrum while others ($\rm B_1 \sim B_4$ and $\rm C$)
  correspond to the quasi-degenerate neutrino mass spectrum.
In this paper, we concentrate only the hierarchical case of the textures 
$\rm A_1$ and $\rm A_2$ :

\begin{equation}
{\rm A_1}: \ \  M_{\nu}=
\left ( \matrix{
{\bf 0} & {\bf 0} & \times \cr
{\bf 0} & \times & \times \cr
\times & \times & \times \cr} \right ) \ , \qquad\qquad
{\rm A_2}:\ \ 
 M_{\nu}=
\left ( \matrix{
{\bf 0} & \times & {\bf 0} \cr
\times & \times & \times \cr
{\bf 0} & \times & \times \cr} \right ) \ ,
\label{texture}
\end{equation}
\noindent where the charged lepton mass matrix is the diagonal one.
Since the component  $M_{\nu}(1,1)$  is zero in these textures, 
the amplitude for the neutrinoless double beta decay vanishes to lowest order 
in neutrino masses.

 In principle these textures are given at the low energy scale because
 the experimental data are put to determine zeros, however, the structure
of zeros in the mass matrix is not changed by the one-loop 
renormalization group  equations  \cite{RGE}.
Therefore, we discuss the see-saw realization in these textures
 at the right-handed Majorana neutrino mass scale:
\begin{equation}
\bf M_\nu=  m_D\   M_R^{\rm -1}\  m_D^{\rm T} \ ,
\end{equation} 
\noindent where  $\bf m_D$ and  $\bf M_R$ are the Dirac neutrino mass
matrix and the right-handed Majorana neutrino one, respectively.
As far as we exclude the possibility that  zeros are 
originated from cancellations among coefficients in the see-saw mechanism, 
the see-saw realization of these seven textures are not trivial.
Then, these zeros should come from zeros of the Dirac neutrino mass matrix 
and the right-handed Majorana neutrino one.
 These results are summarized in the reference \cite{KKST}.

Once the right-handed Majorana neutrino mass matrix is specified,
the Dirac neutrino ones are selected in order to
reproduce the textures in eq.(\ref{texture}). The simplest right-handed 
Majorana neutrino mass matrix is the diagonal one, 
but there is no solution to give desirable texture zeros unless we consider
 cancellations between matrix elements of the Dirac and Majorana ones.
Therefore, we take a following  texture
 for the right-handed Majorana neutrinos 
\begin{equation}
{\bf M_R} =\left ( \matrix{  {\bf 0}  & \times &  {\bf 0}\cr
 \times &   \times &  {\bf 0}\cr {\bf 0} & {\bf 0} & \times \cr} \right ) \ ,
\label{MR}
\end{equation}
\noindent 
where $\times's$ denote non-zero entries.
It should be noted that specifying the texture of the right-handed Majorana 
neutrinos is a choice of weak basis.
Furtheremore we can take the matrix  in eq.(\ref{MR}) to be real.  Then,
the CP violating phases appear only in the Dirac neutrino mass matrix.
In this case, we have six Dirac neutrino mass matrices, which reproduce
  the  texture  $\rm A_1$ in eq.(\ref{texture}):
\begin{equation}
{\bf m_D} =\left ( \matrix{   {\bf 0} & \times &  {\bf 0}\cr
  {\bf 0} & \times &  \times\cr  \times & \times  & \times \cr} \right ) \ ,
\quad 
\left ( \matrix{  {\bf 0} & \times &    {\bf 0}\cr
  {\bf 0} & \times &  \times\cr  \times  &{\bf 0} &  \times \cr} \right ) \ ,
\quad 
\left ( \matrix{ {\bf 0} &\times &   {\bf 0}\cr
  {\bf 0} & \times &   \times\cr  \times & \times  & {\bf 0} \cr} \right ) \ ,
\quad 
\left ( \matrix{   {\bf 0} & \times &   {\bf 0}\cr
  {\bf 0} & \times &   \times\cr  \times & {\bf 0} & {\bf 0}\cr} \right ) \ ,
\label{A10}
\end{equation}
\noindent and
\begin{equation}
{\bf m_D} =\left ( \matrix{  {\bf 0} & \times &   {\bf 0}\cr
  {\bf 0} &  {\bf 0} &  \times\cr  \times & \times  & \times \cr} \right ) \ ,
\quad
\left ( \matrix{   {\bf 0} & \times &    {\bf 0}\cr
  {\bf 0} & {\bf 0} & \times\cr \times  & {\bf 0} & \times \cr} \right ) \ ,
\label{A11}
\end{equation}
 \noindent
where $\times's$ denote  complex numbers.
As seen later the textures in eq.(\ref{A11}) give   different  predictions
 from the ones in  eq.(\ref{A10}) for the decay rate of  
$\mu\rightarrow e \gamma$. 

 For the texture $\rm A_2$ in eq.(\ref{texture}), we obtain
\begin{equation}
{\bf m_D} =\left ( \matrix{  {\bf 0} & \times &    {\bf 0}\cr
 \times &   \times &  \times\cr   {\bf 0}  & \times &\times \cr} \right ) \ ,
\quad 
\left ( \matrix{  {\bf 0} & \times &    {\bf 0}\cr
 \times &   \times &  \times\cr  {\bf 0} & {\bf 0} & \times \cr} \right ) \ ,
\quad 
\left ( \matrix{ {\bf 0} & \times &     {\bf 0}\cr
 \times &  \times &{\bf 0} \cr   {\bf 0} & \times & \times \cr} \right ) \ ,
\label{A20}
\end{equation}
\noindent and
\begin{equation}
{\bf m_D} =\left ( \matrix{  {\bf 0} &\times &    {\bf 0}\cr
  \times &  {\bf 0} &\times\cr  {\bf 0}  & \times & \times \cr} \right ) \ ,
\quad
\left ( \matrix{  {\bf 0} & \times &    {\bf 0}\cr
  \times & {\bf 0} & \times\cr {\bf 0} & {\bf 0}  & \times \cr} \right ) \ ,
\quad
\left ( \matrix{  {\bf 0} & \times &   {\bf 0}\cr
 \times & {\bf 0} & {\bf 0}\cr  {\bf 0}&\times & \times \cr}\right ) \  .
\label{A21}
\end{equation}
The textures in eq.(\ref{A21}) also  give  different  predictions
 from the ones in  eq.(\ref{A20}) for $\mu\rightarrow e \gamma$. 

Although there are phases in  all non-zero entries, 
 those phases  are not necessarily physical quantities.
 Actually, three phases are removed 
by the re-definition of the left-handed neutrino fields.
There is no freedom of the  re-definition for the right-handed ones
in the  basis with  real $\bf M_R$.
As seen in eqs.(\ref{A10}), (\ref{A11}), (\ref{A20}), (\ref{A21}),
 the number of independent phases  are classified as:
  one   texture  of three zeros with three phases,  
  three textures of four  zeros with two  phases and 
  two   textures of five  zeros with one phase. 
Thus, we get the reduction of the number of phases in our textures while 
there are six independent phases in the neutrino mass matrix without zeros.
 
%%%%%%%%%%%%%%%%%%%%%%%%%%%%%%%%%%%%%%%%%%%
%%%%%%%%%%%%%% mu -> e + gamma %%%%%%
%%%%%%%%%%%%%%%%%%%%%%%%%%%%%%%%%%%%%
 What kind of experiments can test these textures? 
It is well known that the Yukawa coupling of the neutrino contributes to
 the lepton flavor violation (LFV).  
Many authors have studied the LFV in the minimal supersymmetric standard 
model (MSSM) with right-handed neutrinos assuming the relevant neutrino 
mass matrix \cite{Borz}$\sim$\cite{our}.
In the MSSM with soft breaking terms, 
there exist  lepton flavor violating terms such as 
off-diagonal elements of slepton mass matrices 
and trilinear couplings (A-term).
It is noticed that large neutrino Yukawa couplings and large lepton mixings 
 generate the large LFV in the left-handed slepton masses. For example, 
the decay rate of  $\mu\rightarrow e \gamma$ can be approximated as follows:

\begin{eqnarray}
{\rm \Gamma} (\mu\rightarrow e \gamma) \simeq 
\frac{e^2}{16 \pi} m^5_{\mu} F
 \left| \frac{(6+2 a_0^2)m_{S0}^2}{16 \pi^2}
({\bf Y_\nu  Y_\nu^\dagger})_{21} \ln\frac{M_X}{M_R}  \right|^2\ ,
\label{rate}
\end{eqnarray}
\noindent where the neutrino Yukawa coupling matrix  $\bf Y_\nu$ 
 is given as  ${\bf Y_\nu}={\bf m_D}/v_2$ ($v_2$ is a VEV of Higgs)
at the right-handed mass scale $M_R$, and 
  $F$ is a function of masses and mixings for SUSY particles.  
In eq.(\ref{rate}),
 we assume the universal scalar mass $(m_{S0})$ for all scalars and
the universal A-term $(A_f=a_0 m_{S0} Y_f)$ at the GUT scale $M_X$. 
Therefore  the branching ratio $\mu \rightarrow e \gamma$ depends 
considerably on the texture of the Dirac neutrino \cite{Sato,Casas,our}.

The magnitude of $({\bf m_D m_D^\dagger})_{21}$ is important
to predict the branching ratio of $\mu\rightarrow e \gamma$
\footnote{Although the right-handed Majorana neutrino mass matrix is not 
a diagonal one in our case,  ${\bf m_D m_D^\dagger}$ does not depend 
on the right-handed sector.}
.
Many works have shown that this branching ratio is too large 
\cite{Sato,Casas}.
However, zeros in the Dirac neutrino mass matrix may lead to  
$({\bf m_D m_D^\dagger})_{21}=0$, and then it suppress the
 $\mu\rightarrow e \gamma$ decay.
 The branching ratio is safely predicted to be below the present experimental
 upper bound $1.2\times 10^{-11}$ \cite{exp} due to zeros.
Actually, the case of $({\bf m_D m_D^\dagger})_{21}=0$ was examined
carefully in ref. \cite{Ellis}.
We have  several textures, which lead to $({\bf m_D m_D^\dagger})_{21}=0$
as seen in eqs.(\ref{A11}) and (\ref{A21}).
 Therefore  we examine the textures of  the Dirac neutrinos in
eqs.(\ref{A11}) and (\ref{A21})  for  the leptogenesis in the following.

%%%%%%%%%%%%%%%%%%%%%%%%%%%%%%%%%%%%%%%%%%%%%%%%%%%%%%%%%%%%%%%%%%
%%%%%%%%%%%%%%%%  A1 Case  %%%%%%%%%%%%%%%%%%%%%%%%%%%%%%%%%%%%%%%
%%%%%%%%%%%%%%%%%%%%%%%%%%%%%%%%%%%%%%%%%%%%%%%%%%%%%%%%%%%%%%%%%%
 At first, let us  discuss the texture in the $\rm A_1$ case 
 in eq.(\ref{texture}).
Key ingredients are the structure of CP violating phases
and the magnitudes of each entry for the Dirac neutrino mass matrices 
in  eq.(\ref{A11})  
since the righ-handed Majorana neutrino mass matrix is taken to be real.
 Although the non-zero entries $\times$ in the Dirac neutrino mass matrix
are  complex, three phases are removed 
by the re-definition of the left-handed neutrino fields.
There is no freedom of re-definition for the right-handed ones
in the basis with  real $\bf M_R$.
Furthermore, we move to the  diagonal basis of the right-handed Majorana 
neutrino mass matrix in order to calculate the magnitude of the leptogenesis.
Then,  the Dirac neutrino mass matrices $\bf \bar m_D$ in the new basis
is given  as follows:
\begin{equation}
  {\bf \bar m_D} =  {\bf P_L \  m_D \ O_R} \ ,
\end{equation}
\noindent
where $\bf P_L$ is a diagonal phase matrix and $\bf O_R$ is the 
orthogonal matrix which diagonalizes $\bf M_R$ as  $\bf O_R^T M_R O_R$
in eq.(\ref{MR}).
Since  the phase matrix $\bf P_L$ can remove one phase in each row of
 $\bf m_D$, three phases disappear  in ${\bf \bar m_D}$.

 By taking three eigenvalues of $\bf M_R$  as follows 
\footnote{The minus sign of $M_1$ is necessary to reproduce 
$\bf M_{R}$  in eq.(\ref{MR}).  This minus sign is transfered to  
$\bf  m_D$ by the right-handed diagonal phase matrix $diag(i,1,1)$.}:
\begin{equation}
  M_1=-\l^m  M_0  \  , \qquad M_2= \l^n  M_0  \ , \qquad  M_3=  M_0  \ ,  
\end{equation}
\noindent  where $\l\simeq 0.22$, 
 we obtain the orthogonal matrix $\bf O_R$ as 
\begin{equation}
  {\bf O_R} =  \left ( \matrix{ \cos{\theta} &  \sin{\theta}&  0\cr 
 - \sin{\theta} &  \cos{\theta} & 0 \cr  0 &  0  & 1 } \right )  \ ,
\qquad\quad  \tan^2 {\theta}= \l^{m-n} \ .
\end{equation}

We take  $m=4,\ n=2$  as a typical case
in order to  find links between the leptogenesis and 
the low energy CP violation.
 The magnitude of each entry in $M_{\nu}$ ($\rm A_1$) is given
to be consistent with the experimental data as follows:
\begin{equation}
 M_{\nu}=
m_0\left ( \matrix{ 0 &  0 & \l \cr  0 & 1& 1\cr \l & 1 & 1\cr} \right ) \ ,
\end{equation}
\noindent where $m_0\simeq \sqrt{\Delta m^2_{\rm atm}}/2$ with
$\Delta m^2_{\rm atm}\simeq 2.5\times 10^{-3}{\rm eV^2}$ is taken.
  Then two mass matrices  ${\bf  m_D}$  in eq.(\ref{A11}) 
can be  parametrized in the new basis as follows:
\begin{equation}
{\bf \bar m_D} =m_{D0}\left ( \matrix{ 0 & a \l^2 &  0\cr
  0 &  0 &  b \cr  c \l^2 e^{i \rho} &  d \l^k  e^{i \sigma}& f \cr} 
 \right ) O_R  , \  (k\geq 1)\  {\rm (A11)} ;
\ \ 
 m_{D0}\left ( \matrix{ 0 & a \l^2 &  0\cr
  0 &  0 &  b \cr  c \l^2 e^{i \rho} &  0 & f \cr} 
 \right )  O_R  ,\    {\rm (A12)} 
\label{DA1}
\end{equation}
 \noindent
where  $k$ is a positive integer with $k\geq 1$,
 and $a$, $b$, $c$, $d$, $f$ are real and order one coefficients.
The magnitude of $m_{D0}$ is determined by the relation
 $m_{D0}^2 \simeq m_0 M_0$.
 There are two independent phases in  (A11), but
only one phase in (A12) as seen in  eq.(\ref{DA1}).
 The texture in (A12) corresponds to the $d=0$ limit in (A11).

In the limit $M_1 \ll M_2,\ M_3$, the lepton number
asymmetry (CP asymmetry) for the lightest heavy Majorana neutrino ($N_1$) 
decays into $l^{\mp} \phi^{\pm}$ \cite{ep} is given by

\begin{equation}
\epsilon_1 =
\frac{\Gamma_1-\overline{\Gamma_1}}{\Gamma_1+\overline{\Gamma_1}}
\simeq  -\frac{3}{16\pi v^2} \left (
  \frac{{\rm Im}[\{({\bar m_D}^{\dagger} \bar m_D)_{12}\}^2]}
{({\bar m_D}^{\dagger} \bar m_D)_{11}}\frac{M_1}{M_2}  +
  \frac{{\rm Im}[\{({\bar m_D}^{\dagger} \bar m_D)_{13}\}^2]}
{({\bar m_D}^{\dagger} \bar m_D)_{11}}\frac{M_1}{M_3}\right )\ ,
\label{epsilon} 
\end{equation}
\noindent where $v=174$ GeV.
The lepton asymmetry $Y_L$ is related to the CP asymmetry through the relation
\begin{equation} 
Y_L=\frac{n_L-n_{\bar{L}}}{s}=\k\,\frac{\epsilon_1}{g_{\ast}}\ ,
\end{equation}
where $s$ denotes the entropy density, 
$g_{\ast}$ is the effective number of relativistic degrees of freedom
contributing to the entropy and $\kappa$ is 
the so-called dilution factor which
accounts for the washout processes (inverse decay and lepton number violating
scattering). 
In the case of the standard model (SM), one gets $g_{\ast}=106.75$.

The produced lepton asymmetry $Y_L$ is converted into a net baryon asymmetry
$Y_B$ through the $(B+L)$-violating sphaleron processes. One  finds the
relation \cite{harvey}
\begin{equation} 
Y_B=\xi \ Y_{B-L}=\frac{\xi}{\xi-1} \ Y_L \  , \qquad
\xi= \frac{8\ N_f + 4\ N_H}{22\ N_f + 13\ N_H} \ ,
\end{equation}
where $N_f$ and $N_H$ are the number of fermion families and complex Higgs
doublets, respectively. Taking into account  $N_f=3$ and $N_H=1$ in the SM, 
we get 
\begin{equation}  
Y_B \simeq -\frac{28}{51}\,Y_L \ .
\end{equation}

On the other hand, the low energy CP violation, which is a measurable 
quantity in the long baseline neutrino oscillations \cite{CP}, 
is given by the Jarlskog determinant $J$ \cite{J}, which is calculated by
\begin{equation}  
   det [{\bf  M_\ell M_\ell^\dagger,  M_\nu M_\nu^\dagger}]
= -2 i   J  (m_\tau^2-m_\mu^2) (m_\mu^2-m_e^2)(m_e^2-m_\tau^2)
 (m_3^2-m_2^2) (m_2^2-m_1^2)(m_1^2-m_3^2)  ,
\end{equation}
\noindent where $\bf M_\ell$ is the diagonal charged lepton mass matrix, 
and $m_1$, $m_2$, $m_3$ are neutrino masses.

 It is very interesting to investigate links
between the leptogenesis ($\epsilon_1$) and the low energy CP violation ($J$) 
in the textures of eq.(\ref{DA1}).
In  the texture (A11), $\epsilon_1$ and $J$ are given as
\begin{eqnarray}  
 \epsilon_1 &\simeq&  -\frac{3 m_{D0}^2}{16\pi v^2}\  [
  d^2\l^{2(k+1)} \frac{c^2\sin{2(\rho-\sigma)}-2c d\l^{k-1}
    \sin{(\rho-\sigma)}}  
  {c^2+d^2 \l^{2(k-1)}-2 c d\l^{k-1}\cos{(\rho-\sigma)}} \nonumber \\
&+&  f^2 \l^4 \frac{c^2\sin{2\rho}+d^2 \l^{2(k-1)}\sin{2\sigma}
 -2 c d \l^{k-1}\sin{(\rho+\sigma)}}
           {c^2+d^2 \l^{2(k-1)}-2 c d\l^{k-1}\cos{(\rho-\sigma)}} ] \  , 
 \nonumber\\
 && \hskip 11 cm (k\geq 1)\\
J &\simeq&   \frac{1}{64}a^2 b^4 c^3 f^2  \l^2
 \ [\  c \sin 2\rho - 2 d \l^{k-1}\sin ({\rho+\sigma}) ]
  \ \frac {\Delta m_{\rm atm}^2}{\Delta m_{\rm sol}^2}\ , 
\nonumber
\end{eqnarray}
\noindent respectively.
Both  $\epsilon_1$ and $J$ are  given in terms of  
two  phases $\rho$ and $\sigma$.
 However, the dominat terms are given  by phase $\rho$  
 as far as  $k\geq 2$.   It is remarked  that the leptogenesis is linked 
strongly with  the CP violation in the neutrino oscillations.
 The relative sign of $\epsilon_1$ and $J$ is  opposite in this case.
  The case of $k= 1$ is exceptional one. 
Detailed  numerical study will be  needed  in  $k= 1$ 
because both  $\epsilon_1$ and $J$ depend on 
parameters $c$, $d$, $\rho$ and $\sigma$.

The case of the texture (A12) is clear because there is only one phase $\rho$.
 In this case,   $\epsilon_1$ and $J$ are given
in the  $d=0$ limit of the texture (A11) as follows:
\begin{eqnarray}  
&&\epsilon_1 \simeq -\frac{3 m_{D0}^2}{16\pi v^2}\ f^2 \ \l^4\ \sin 2\rho\ , 
\nonumber\\
\label{RA12}\\
&&J\simeq   \frac{1}{64}a^2 b^4 c^4 f^2  \l^2\
  \frac {\Delta m_{\rm atm}^2}{\Delta m_{\rm sol}^2}\   \sin 2\rho \ .
\nonumber
\end{eqnarray}
\noindent
The relative sign of $\epsilon_1$ and $J$ is  also opposite.
It may be helpful to note that only the second term in the right-handed side 
of eq.(\ref{epsilon})  contributes  on  $\epsilon_1$ 
since ${\rm Im}[\{({\bar m_D}^{\dagger} \bar m_D)_{12}\}^2]$ vanishes.

On the other hand,  taking the LMA-MSW solution 
${\Delta m_{\rm sol}^2}/{\Delta m_{\rm atm}^2}= \l^3\sim \l^2$ and
$\sin 2\rho\simeq 1$,
we predict  $J \simeq 0.01$, which is rather large and then is favored
 for the experimental measurement.
%%%%%%%%%%%%%%%%%%%%%%%%%%%%%%%%%%%%%%%%%%%%%%%%%%%
%%%%%%%%%%%%%%%% A2 Case  %%%%%%%%%%%%%%%%%%%%%%%%%
%%%%%%%%%%%%%%%%%%%%%%%%%%%%%%%%%%%%%%%%%%%%%%%%%%%

Next, consider the case $\rm A_2$  for $M_\nu$.
Since  the magnitude of each entry in $M_{\nu}$ is given
to be consistent with the experimental data as follows:
\begin{equation}
 M_{\nu}=
 m_0 \left ( \matrix{ 0 & \l & 0 \cr \l & 1& 1\cr 0 & 1 & 1\cr} \right ) \ ,
\end{equation}
\noindent 
three  mass matrices  ${\bf  m_D}$  in eq.(\ref{A21}) can be parametrized 
in the new basis as follows:
\begin{equation}
{\bf \bar m_D} = m_{D0}\left ( \matrix{   0 & a \l^2 &   0\cr
   b\l^2  & 0 &   c  e^{i \rho} \cr    0 & d \l^p e^{i \sigma} & f \cr} 
 \right ) O_R    , \   {\rm (A21)}\ ;
\quad
m_{D0}\left ( \matrix{ 0 & a \l^2 &   0\cr
   b\l^2  & 0 &   c  \l^q e^{i \rho} \cr    0 & d \l  e^{i \sigma} & f \cr} 
 \right ) O_R   , \   {\rm (A22)}
\label{DA2}
\end{equation}
\noindent where $p\geq 1$  and $q\geq 0$, and 
\begin{equation}
  {\bf \bar m_D}= m_{D0} \left ( \matrix{   0 & a \l^2 &   0\cr
   b\l^2  &0 &   c  e^{i \rho} \cr    0 & 0  & f \cr} 
 \right )  O_R \ , \ \  {\rm (A23)}\ ;
\quad
    m_{D0}\left ( \matrix{ 0 & a \l^2 &   0\cr
   b\l^2  & 0 &   0 \cr    0 & d \l  e^{i \sigma} & f \cr} 
 \right ) O_R \   , \ \   {\rm (A24)}\ .
\label{DA20}
\end{equation}
 \noindent
 The textures in (A23) and (A24) correspond to the $d=0$ and $c=0$ limits 
of  (A21) and (A22) in eq.(\ref{DA2}), respectively.

For  the texture (A21) $\epsilon_1$ and $J$ are given as
\begin{eqnarray}  
&& \epsilon_1 \simeq  \frac{3 m_{D0}^2}{16\pi v^2} \ 
\frac{1}{b^2+d^2\l^{2(p-1)}}  \l^4  
    [b^2 c^2\sin 2\rho -d^2 f^2\l^{2(p-1)} \sin 2\sigma-
   2 bcdf\l^{p-1} \sin(\rho-\sigma)] \    ,  \nonumber \\
     && \hskip 12.5 cm (p\geq 1)  \label{RA21} \\
&& J  \simeq   \frac{1}{64}a^2 b^3 f^2 \l^2\
   \frac {\Delta m_{\rm atm}^2}{\Delta m_{\rm sol}^2}\  
[ \  b c^2 f^2 \sin 2\rho \nonumber \\
&&+d\l^{p-1} \{ 2c^3f  \sin(\rho-\sigma)+
 bc^2d\l^{p-1}  
\sin 2(\rho-\sigma)+b^3d\l^{p-1} \sin 2\sigma+2b^2cf\sin(\rho+\sigma) \}]  ,
\nonumber
\end{eqnarray}
\noindent and for  the texture (A22) $\epsilon_1$ and $J$ are given as
\begin{eqnarray}  
 &&\epsilon_1 \simeq  \frac{3 m_{D0}^2}{16\pi v^2} \ 
\frac{1}{b^2+d^2} \ \l^4 \ 
    [b^2 c^2\l^{2q}\sin 2\rho -d^2 f^2 \sin 2\sigma-2 bcdf\l^{q}
 \sin(\rho-\sigma)] \    , \nonumber\\
  && \hskip 12.5 cm (q\geq 0) \label{RA22}\\
&&J  \simeq   \frac{1}{64} a^2 b^3 f^2  \l^2\
   \frac {\Delta m_{\rm atm}^2}{\Delta m_{\rm sol}^2}\  
[ \ b^3 d^2 \sin 2\sigma \nonumber \\
&&+c\l^{q} \{2c^2 df \l^{2q} \sin(\rho-\sigma)+
 bcd^2\l^{q}  \sin 2(\rho-\sigma)+bcf^2\l^{q} \sin 2\rho+2b^2df\sin(\rho+\sigma) \}]  .   \nonumber
\end{eqnarray}
\noindent
 Both   $\epsilon_1$ and $J$ are dominated by one phase
 as far as $p\geq 2$ and  $q\geq 1$. Therefore, the leptogenesis is linked 
strongly with  the CP violation in the neutrino oscillations
 except for the case of $p=1$ and  $q=0$.

%%%%%%%%%%%%%%%%%%%%%%%%%%%%%%%%%%%%%%%%%%%%%%%%%%%%%%%%%%%%%%%%%%%%%%%
In the case of the textures (A23) and (A24),  $\epsilon_1$ and $J$ are given
 by putting  $d=0$ and $c=0$ in eqs.(\ref{RA21}) and  (\ref{RA22}), 
respectively.  Therefore
 $\epsilon_1$ and $J$ are simply given as
\begin{eqnarray}  
&& \epsilon_1 \simeq  \frac{3 m_{D0}^2}{16\pi v^2} \ 
c^2 \ \l^4 \ \sin 2\rho \    , \nonumber\\
\label{RA23}\\
&&J\simeq     \frac{1}{64}a^2 b^4 c^2 f^4 \l^2\
   \frac {\Delta m_{\rm atm}^2}{\Delta m_{\rm sol}^2}\   \sin 2\rho \ ,
\nonumber
\end{eqnarray}
\noindent for (A23), and 
\begin{eqnarray}  
&& \epsilon_1 \simeq  -\frac{3 m_{D0}^2}{16\pi v^2} \ 
\frac{d^2 f^2}{b^2+d^2} \ \l^4 \ \sin 2\sigma \    , \nonumber\\
\label{RA24}\\
&&J\simeq     \frac{1}{64}a^2 b^6 d^2 f^2 \l^2\
   \frac {\Delta m_{\rm atm}^2}{\Delta m_{\rm sol}^2}\   \sin 2\sigma\ ,
\nonumber
\end{eqnarray}
\noindent for (A24), respectively.
The relative sign of $\epsilon_1$ and $J$ is same for (A23) while opposite 
for (A24) as seen in  eqs.(\ref{RA23}) and (\ref{RA24}).
Therefore, these textures are  testable in the future.

%%%%%%%%%%%%%%%%%%%%%%%%%%%%%%%%%%%%%%%%%%%%%%%%%%%%%%
%%%%%%%%%%%%%%%%  Leptogenesis  %%%%%%%%%%%%%%%%%%%%%% 
%%%%%%%%%%%%%%%%%%%%%%%%%%%%%%%%%%%%%%%%%%%%%%%%%%%%%%
 In order to calculate the baryon asymmetry, 
we need the dilution factor involves the integration of the 
full set of Boltzmann equations \cite{pu}. 
A simple approximated solution which has been
frequently used is given by \cite{kolb}
\begin{eqnarray}
%\kappa =\left\{
%\begin{array}{l}  
%\sqrt{\,0.1\,K_R}\,\exp\left(-\frac{4}{3}\sqrt[4]{\,0.1\,K_R}\right)
%\quad , \quad  K_R \leq  10^{6}  \\
%\\
% 0.24 (K_R\,\ln K_R)^{-3/5}\quad , \quad
%10 \leq K_R  \leq 10^6 \\
%\\
%1/(2 K_R) \quad , \quad 1 \leq K_R \leq 10 \\
%\\
%1  \quad , \quad 0  \leq K_R  \leq 1
%\end{array}
%\right.
\kappa=\frac{0.3}{K_R} (\ln K_R)^{-0.6}\ , \qquad (10 \leq K_R  \leq 10^6) \ ,
\end{eqnarray}
\noindent
where the parameter $K_R$ 
is defined as the ratio of the thermal average of the $N_1$ decay
rate and the Hubble parameter at the temperature $T = M_1\ $,
\begin{eqnarray} 
K_R=\frac{M_P}{1.7 \times 8 \pi v^2  \sqrt{g_{\ast}}}
\frac{({\bar m_D}^{\dag}\,\bar m_D)_{11}}{M_1}\ ,
\end{eqnarray}
where $M_P \simeq 1.22 \times 10^{19}$~GeV is the Planck mass.
The value of $K_R$ is approximately $\sim 23$,  which is independent of
$M_1$, in our textures.  
Therefore, the dilution factor is not so suppressed.

%%%%%%%%%%%%%%%%%%%%%%
%%%%%%%%   fig.1.
%%%%%%%%%%%%%%%%%%%%%%
\begin{figure}
\epsfxsize=10.0 cm
\centerline{\epsfbox{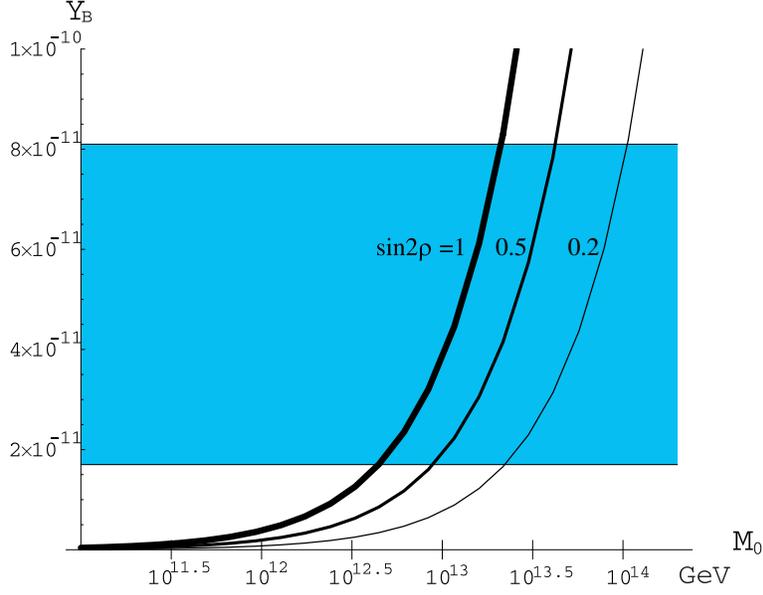}}
\caption{Predicted $Y_{B}$ versus $M_{0}$ for $\sin 2\rho=0.2,\ 0.5,\ 1$ in 
the texture A12. The gray region is allowed by the observed baryon asymmetry.}
\end{figure}
%%%%%%%%%%%%%%%%%%%%%%
%%%%%%%%%%%%%%%%%%%%%%
By using this approximate dilution factor,
 we have calculated $Y_B$ in our textures.
Since $\epsilon_1\simeq \l^4$ in all cases of our textures,
we calculate $Y_B$ for the  case of (A12) with  $f=1$ in eq.(\ref{RA12})
as a typical one.
Our numerical result is shown in figure 1.
As seen in figure 1, the predicted  $Y_B$ is consistent with
the observed value $Y_B=(1.7 \sim  8.1)\times 10^{-11}$
for  $M_0=10^{13}\sim 10^{14}\G$ and $\sin 2\rho=0.2\sim 1$.
More detailed analyses will be given elsewhere.
Thus, our textures of the Dirac neutrinos is favored
in the scenario of the leptogenesis.

%%%%%%%%%%%%%%%%%%%%%%%%%%%%%%%%%%%%%%%%%%%%%%%%%%%%%%
%%%%%%%%%%%%%%%%  Summary  %%%%%%%%%%%%%%%%%%%%%%%%%%% 
%%%%%%%%%%%%%%%%%%%%%%%%%%%%%%%%%%%%%%%%%%%%%%%%%%%%%%
 The discussion and summary are  given as follows.

We have studied  the leptogenesis in the neutrino mass matrix with two zeros, 
which reduces the number of CP violating phases. After see-saw realization 
in the textures $\rm A_1$ and $\rm A_2$, we have found  that
 one texture with three zeros has three phases, three textures
  with four zeros have two phases and two textures with five zeros have
 only one phase for the Dirac neutrino mass matrices, respectively.
Taking a typical mass hierarchy of the right-handed Majorana neutrinos,
we have calculated the relevant quantities  
for the leptogenesis and the low energy CP violation,  $\e_1$ and $J$,
 respectively. The leptogenesis is linked 
strongly with  the CP violation in the neutrino oscillations
in our textures although there are a few exceptional cases.
 We have also given the prediction  $Y_B$, 
which is consistent with the observed value.
We will present  more systematic and detailed analyses elsewhere.
In order to test the link between the leptogenesis and the low energy 
CP violation, we expect that the measurement of the CP violation 
in the experiments of the long baseline neutrino oscillations.

\vskip 1  cm
 
We are grateful to T. Yanagida and P. H. Frampton
 for important comments on the number of phases in textures.
 We also thank  T. Endoh,  S. K. Kang and T. Morozumi  
for useful discussions  at the YITP-W-02-05 on 
"Flavor mixing, CP violation and Origin of matter".
This research is supported by the Grant-in-Aid for Science Research,
 Ministry of Education, Science and Culture, Japan (No.12047220). 

\newpage
%%%%%%%%%%%%%%%%%%%%%%%%%%%%%%%%%%%%%%%%%%%%%%%%%%%%%%%%%%%%%%%%%%%%%%%%%%%%%%%%%%%%%%%%

%%%%%%%%%%%%%%%%%%%%%%%%%%%%%%%%%%%%%%%%%%%%%%
\end{document}